# Gallium Bismuth Halides GaBi-$X_2$ (X= I, Br, Cl) Monolayers with Distorted Hexagonal Framework: Novel Room-Temperature Quantum Spin Hall Insulators

Linyang Li[1], Ortwin Leenaerts[1], Xiangru Kong[2], Xin Chen[3], Mingwen Zhao[3] and François M. Peeters[1]

[1] *Department of Physics, University of Antwerp, Groenenborgerlaan 171, B-2020 Antwerp, Belgium*

[2] *International Center for Quantum Materials, Peking University, 100871 Beijing, China*

[3] *School of Physics and State Key Laboratory of Crystal Materials, Shandong University, 250100 Jinan, China*

**Abstract:** Quantum Spin Hall (QSH) insulators with a large topologically nontrivial bulk gap are crucial for future applications of the QSH effect. Among these, group III-V monolayers and their halides with chair structure (regular hexagonal framework, RHF) were widely studied. Using first-principles calculations, we propose a new structure model for the functionalized group III-V monolayers, which consist of rectangular GaBi-$X_2$ (X=I, Br, Cl) monolayers with a distorted hexagonal framework (DHF). These structures have a much lower energy than the GaBi-$X_2$ monolayers with chair structure. Remarkably, the DHF GaBi-$X_2$ monolayers are all QSH insulators, which exhibit sizeable nontrivial band gaps ranging from 0.17 eV to 0.39 eV. Those band gaps can be widely tuned by applying different spin-orbit coupling (SOC) strengths, resulting in a distorted Dirac cone.

**Keywords:** QSH insulators, first-principles calculations, GaBi-$X_2$ (X=I, Br, Cl) monolayers, distorted hexagonal framework, distorted Dirac cone

E-mail:
(L.L.) linyang.li@uantwerpen.be
(O.L.) ortwin.leenaerts@uantwerpen.be
(X.K.) kongxru@pku.edu.cn
(F.M.P.) francois.peeters@uantwerpen.be

**Introduction**

Monolayer graphene was first realized in 2004 and since then two-dimensional (2D) materials play a crucial role in the field of nanomaterials [1]. In the meantime, more and more 2D materials with unusual electronic and spintronic properties have been synthesized, that are promising for applications in quantum devices. For example, the group IV element monolayers, silicene [2-13], germanene [14-17], and stanene [18] were all successfully synthesized on different substrates. Besides graphene and its analogs, layered transition-metal dichalcogenides (TMDs) are another family of 2D materials that have been realized experimentally [19-21]. All these 2D materials can be regarded as being composed of a regular hexagonal framework (RHF), which is usually called the chair structure. Most predictions of new 2D materials are also based on the RHF model, such as the silicon-germanium monolayer [22], HgSe/HgTe monolayer [23], III-Bi monolayer [24], and organic metal frameworks [25-28]. A natural question arises: Can we find other stable structure models for those 2D materials?

Quantum spin Hall (QSH) insulators are another important set of 2D materials [29-31]. The QSH effect has been observed in HgTe/CdTe and InAs/GaSb quantum wells [32-34], but the operating temperature is limited due to their small bulk gap arising from weak spin-orbit coupling (SOC). The necessary low temperature also limits its applications. For the synthesis of the QSH insulators with a large gap, Bi (111) bilayer has been realized in experiment and its time-reversal symmetry-protected edge states have been observed, but there is still experimental discrepancy regarding its topological nature [35-37]. Therefore, searching for more new QSH insulators with a large bulk gap will provide more choices for experiment and application. Starting from graphene [38], many kinds of QSH insulators have been predicted, including graphene with a sandwich structure [39, 40], silicene (germanene/stanene) with RHF [41, 42], stanene with a dumbbell structure [43-45], $MoS_2$ allotropes [46-52], and monolayers containing heavy metal atoms (Bi/Sb/Hf) [24, 53-56]. In the case of particular substrates [57], applying an external field [58], or using hydrogenation/functionalization [59-78], the size of those nontrivial bulk gaps can be further increased. Currently, the largest nontrivial bulk gap is 1.08eV found in $Bi_2F_2$ monolayer [59, 60], which shows that chemical functionalization is a very powerful way to obtain a large nontrivial bulk gap. Using first-principles calculations, some group III-V monolayers were predicted to be QSH insulators with a large bulk gap [24] and it was found that this gap can be enlarged via hydrogenation/functionalization [71-78]. After functionalization, their structures can all be regarded as being composed of RHF. However, similar as the different ways of functionalization of graphene, one may ask if there are other ways of functionalization of group III-V monolayers that simultaneously retain their topologically nontrivial properties?

In order to tackle the above questions, we propose a new functionalization model, called distorted hexagonal framework (DHF), for iodination, bromination, and chlorination of gallium bismuth monolayer (GaBi-$X_2$, X=I, Br, Cl). We find that the ground state energy of the DHF GaBi-$X_2$ monolayers is less than that of the RHF GaBi-$X_2$ monolayers, which were widely studied in previous works [71-78]. Using first-principles calculations, we investigated systematically the structure, stability and electronic property of the DHF GaBi-$X_2$ monolayers. The GaBi framework is robust against different ways of functionalization for the three DHF GaBi-$X_2$ monolayers. Phonon spectra provide convincing evidence for the thermal and dynamical stabilities of those DHF GaBi-$X_2$ monolayers. Remarkably, we find a distorted Dirac cone in the band structure without SOC of the DHF GaBi-$I_2$ monolayer and a large nontrivial bulk gap (0.39 eV) in the band structure with SOC, which is large enough to achieve the QSH effect at room-temperature. Both GaBi-Br$_2$ and GaBi-Cl$_2$ have indirect band gaps without and with SOC. However, we find that a distorted Dirac cone appears when changing the strength of the SOC. Furthermore, the topologically nontrivial property of the bulk gaps is confirmed by the nonzero Z$_2$ topological invariant and the appearance of gapless Dirac state from the edges.

**Calculation Method**

Our first-principles calculations were performed using the plane-wave basis Vienna *ab initio* simulation package (VASP) [79-81]. The electron exchange-correlation functional was treated using the generalized gradient approximation (GGA) in the form proposed by Perdew, Burke, and Ernzerhof (PBE) [82]. The atomic positions and lattice vectors were fully optimized using the conjugate gradient (CG) scheme until the maximum force on each atom was less than 0.01 eV/Å. The energy cutoff of the plane waves was set to 520 eV with an energy precision of $10^{-5}$ eV. For the 2D structure relaxation, the Brillouin zone (BZ) was sampled by using a 11×11×1 Γ-centered Monkhorst-Pack grid, while a 15×15×1 grid was used for the static calculations. A 1×11×1 grid was used for the nanoribbon calculations. The vacuum space was set to at least 15 Å in all the calculations to minimize artificial interactions between neighboring slabs. SOC was included by a second variational procedure on a fully self-consistent basis. The phonon spectra were calculated using a supercell approach within the PHONON code [83].

As our systems preserve time-reversal symmetry and break space-inversion symmetry, we can obtain the Z$_2$ topological invariant by calculating Wannier Charge Centers (WCCs) [84]. In the concept of time-reversal polarization [85], the integer Z$_2$ invariant (Δ) can be written as

$$\Delta = P_\theta(T/2) - P_\theta(0) \bmod 2,$$

where $P_\theta(t)$ is the total charge polarization with the cyclic parameter $t$ and $T$ is the period of cyclic adiabatic evolution. The above equation can be rewritten in terms of WCCs ($\bar{x}_\alpha$):

$$\Delta = \sum_\alpha [\bar{x}_\alpha^I(T/2) - \bar{x}_\alpha^{II}(T/2)] - \sum_\alpha [\bar{x}_\alpha^I(0) - \bar{x}_\alpha^{II}(0)].$$

*I* and *II* indicate the Kramers pairs. In explicit numerical implementations, a more straightforward and more easily automated approach is to track the largest gap in the spectrum of the WCCs. Let $\Delta_m$ be the number of WCCs that appear between neighboring gap centers and $M$ is the total number of changes in $\Delta_m$. The $Z_2$ invariant is then given by:

$$\Delta = \sum_{m=0}^{M} \Delta_m \bmod 2.$$

Therefore, the $Z_2$ invariants can be obtained easily by numerical computations with *ab initio* codes together with the Wannier90 code [86, 87].

**Geometrical Structure**

The three investigated DHF GaBi-$X_2$ monolayers exhibit a similar structure. Figure 1(a) and (b) show the optimized geometrical structure of GaBi-$X_2$. From the top view, the framework of Ga-Bi is DHF, which is totally different from the previous chair models [71-78], which are RHF with three-fold rotation symmetry like silicene/germanene [41, 42], as shown in Figure 1(c) and (d). To obtain the DHF structure, we used a larger rectangular supercell and moved the X atoms from their original location, which is right on top of the Ga/Bi atoms. The fully optimized DHF GaBi-$X_2$ monolayer shows a high symmetric space group of Pca$2_1$. We choose a rectangular primitive cell with lattice constants $a$ along the x direction and $b$ along the y direction. Without loss of generality, we can take the DHF GaBi-$I_2$ monolayer as an example. Its lattice constants are $a = 8.58$ Å and $b = 8.14$ Å. Considering the Ga-Bi bonds, there are three kinds of bonds with different lengths, which is totally different from the RHF GaBi-$I_2$ monolayer. For the direction along y, the length of the Ga-Bi bond is labeled as L1, which is equal to 2.91 Å. For the direction along x, the lengths of the Ga-Bi bonds are labeled as L2 and L3, respectively, as shown in Figure 1(a). L2 is equal to 2.95 Å and L3 is equal to 2.80 Å. L2 and L3 alternate along the x direction. The different lengths of the Ga-Bi bonds lead to the DHF, while the lengths of L1, L2 and L3 are the same in the RHF (Figure 1(c)). For the RHF GaBi-$I_2$ monolayer, all the I atoms are right on the Ga/Bi atoms along the z direction forming a chair configuration (Figure 1(d)), but in the DHF GaBi-$I_2$ monolayer, the bonds of Ga-I and Bi-I are totally different. For Bi-I bonds, one I atom forms a bond with two Bi atoms and the lengths of the two Bi-I bonds can be labeled as M1 and M2. Along the x direction, M1 and M2 alternate, similar to L2 and L3. For Ga-I bonds, labeled as N (Figure 1(b)), the I atom is not right on top of a Ga atom along the z direction, but there is a little deviation between them. For the three DHF GaBi-$X_2$ monolayers, the optimized geometrical structure data are summarized in Table 1. From iodination to chlorination, the lattice constants ($a$ and $b$) diminish slightly and the lengths of M1, M2, and N also decrease, but the lengths of the three Ga-Bi bonds (L1, L2, and L3) remain almost unchanged.

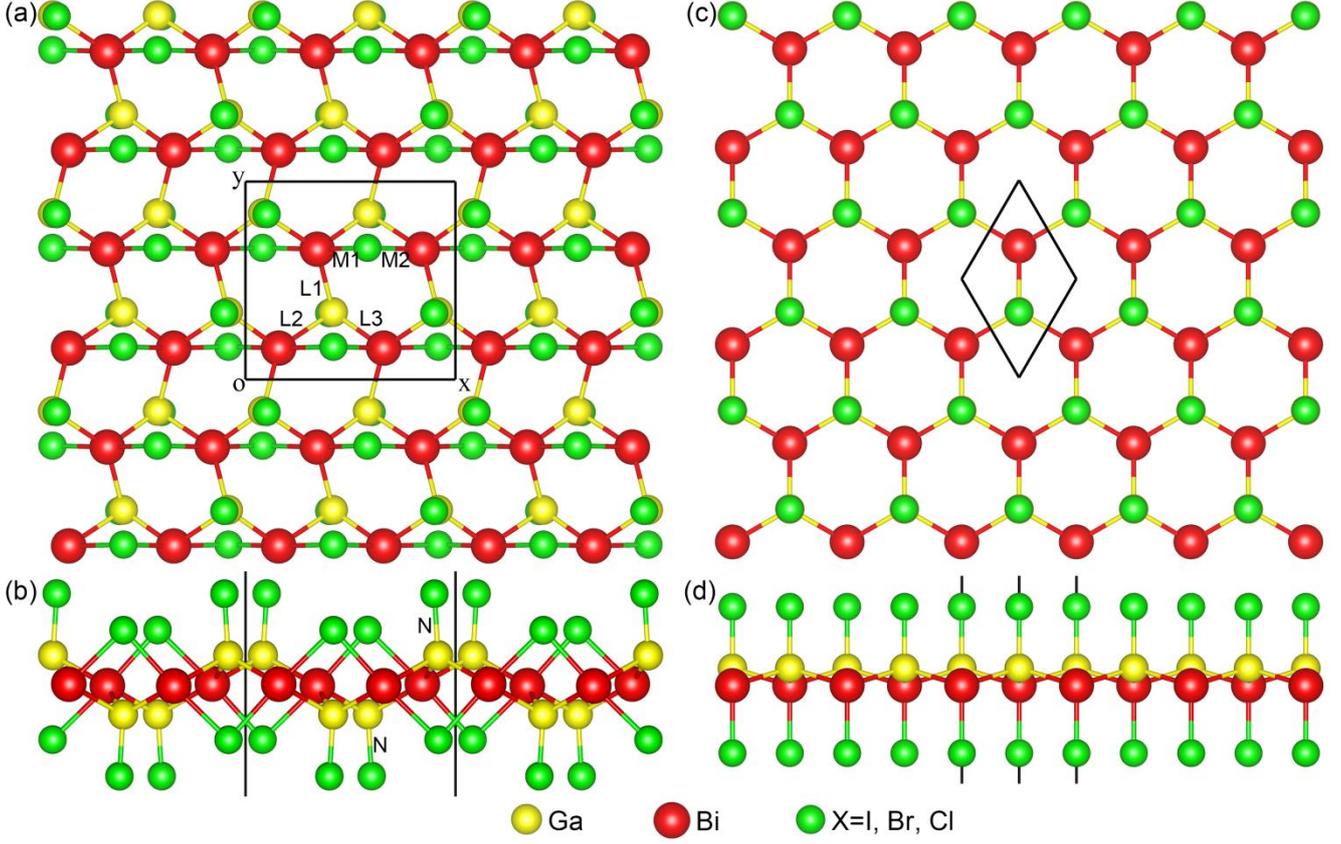

**Figure 1.** Top view (a) and side view (b) of the DHF GaBi-$X_2$. Top view (c) and side view (d) of the RHF GaBi-$X_2$.

**Table 1**. Structure parameters and relative energies of the DHF GaBi-$X_2$ monolayers. $a$ and $b$ are the lattice constants. L1, L2, and L3 are the three lengths of the Ga-Bi bonds. M1 and M2 correspond to the two lengths of the Bi-X bonds and N is the length of the Ga-X bond. The relative formation energy $\Delta E$ is obtained by setting the energy of the RHF GaBi-$X_2$ monolayer as 0. $Z_2$ is the topological invariant.

| | $a$ (Å) | $b$ (Å) | L1(Å) | L2(Å) | L3(Å) | M1(Å) | M2(Å) | N(Å) | $\Delta E$(meV/atom) | $Z_2$ |
|---|---|---|---|---|---|---|---|---|---|---|
| GaBi-$I_2$ | 8.58 | 8.14 | 2.91 | 2.95 | 2.80 | 3.08 | 3.15 | 2.54 | -49 | 1 |
| GaBi-$Br_2$ | 8.36 | 8.03 | 2.89 | 2.95 | 2.79 | 2.88 | 2.95 | 2.32 | -55 | 1 |
| GaBi-$Cl_2$ | 8.22 | 8.00 | 2.89 | 2.95 | 2.79 | 2.73 | 2.79 | 2.17 | -57 | 1 |

**Energy and Stability**

We compare the energies of the RHF GaBi-$X_2$ monolayers, that have been investigated in previous studies [71-78], with those of the newly proposed DHF GaBi-$X_2$ monolayers. We set the energy of RHF GaBi-$X_2$ as zero and then calculated the relative formation energy of DHF GaBi-$X_2$, which we show in Table 1. The DHF GaBi-$X_2$ monolayers all have a substantially lower energy. The RHF GaBi-$X_2$ monolayers exhibit a polarized structure due to the up-down asymmetry, which increases their energies and probably will lead to a curled configuration. In comparison, the DHF GaBi-$X_2$ has no dipole moment. The above two aspects are responsible for the superiority in the structural stability of the DHF GaBi-$X_2$ over the RHF GaBi-$X_2$. To investigate the dynamical stability of DHF GaBi-$X_2$ monolayers, we calculated their phonon spectra and show them in Figure 2. It is clear that the DHF GaBi-$X_2$ monolayers are all free from imaginary frequency modes and are therefore dynamically stable.

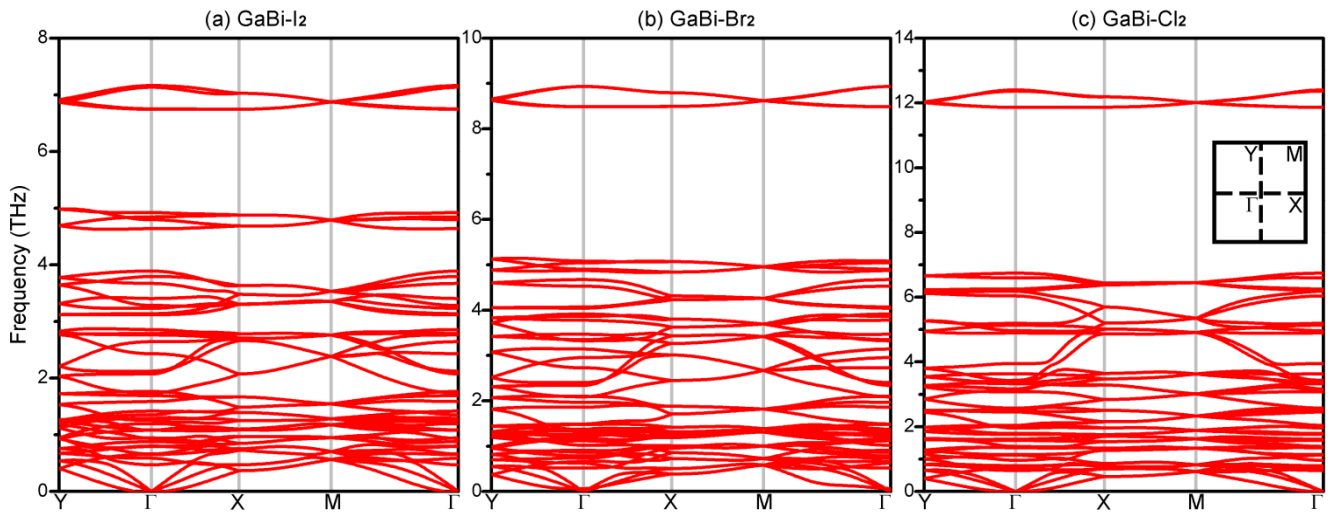

**Figure 2**. Phonon spectra of the DHF GaBi-$X_2$ monolayers along the high-symmetry points in the BZ.

We further confirm the thermodynamic stability of the DHF GaBi-$I_2$ monolayer by first-principles molecular dynamics (MD) simulations. We used a 2×2 supercell to perform MD simulations at 300K

(room-temperature). After 4ps, we show its atomic configuration in Figure S1(a) and (b) (Supplementary Information). It is clear that the DHF structure is preserved from the top view (Figure S1(a)) and the Ga and Bi atoms still keep their three-layer structure with up-down symmetry (Figure S1(b)). Although the I atoms show little deviations compared to their original sites (Figure 1) due to thermal perturbations, neither structure disruption nor structure reconstruction occurs in the DHF GaBi-$I_2$ monolayer, which suggests its stability at room-temperature. It should be noted that although there are some experiments on GaBi(As) [88, 89], InBi [90] and TlBi [91] films, the experimental support for their functionalization is still missing, similar to the experimental absence of other functionalized monolayers (e.g. functionalized Bi and Pb monolayers) [35]. Regarding experimental realization, we suggest to obtain first the GaBi monolayer first and then synthesize the GaBi-$X_2$ monolayers through functionalization. Another possible chemical route is as follows. Taking the DHF GaBi-$I_2$ monolayer as an example, it could be synthetized by: $BiI_3$ [92, 93] + $GaI_3$ [94-96] +4Na → GaBi-$I_2$ (DHF) + 4NaI. Our DFT calculations show that the reaction is exothermic with an energy release of 0.6 eV/atom, which indicates its feasibility for experimental synthesis.

**Electronic Band Structure**

The electronic band structures of the RHF GaBi-$X_2$ monolayers are topologically nontrivial and show a large bulk gap due to the strong SOC at the Γ point [71-78]. The electronic band structures of the DHF GaBi-$I_2$ monolayer obtained from DFT are plotted in Figure 3. For the DHF GaBi-$I_2$ monolayer, the valence and conduction bands meet at a single point along the Γ-X line, giving rise to a distorted Dirac cone, as shown in Figure 3(a). Similar band structures can also be found in other rectangular lattices [97, 98]. The orbital-projected band structure without SOC of the DHF GaBi-$I_2$ monolayer close to the Fermi level is shown in Figure 4(a). It is clearly seen that the distorted Dirac cone mainly comes from the atomic orbitals of the Ga and Bi atoms. For the crossing bands forming the Dirac cone, one band comes from the $p_y$ atomic orbitals (green dots) and the other originates from the $s$, $p_x$, and $p_z$ atomic orbitals (red dots). The band structure with SOC for the DHF GaBi-I$_2$ monolayer shows an indirect band gap (Figure 3(b)). The indirect band gap is 0.39 eV along the Γ-X line, which is less than that of the RHF GaBi-I$_2$ monolayer (0.606 eV) [75]. From its corresponding orbital-projected band structure close to the Fermi level (Figure 4(b)), we find that these bands are mainly due to the $s$ and $p$ atomic orbitals of the Ga and Bi atoms. We now focus on the valence and conduction bands along the Γ-X line. The orbital contributions can be divided into two parts. The first part of the valence band corresponds to the $s$, $p_x$, and $p_z$ orbitals (red dots) and the second part comes from the $p_y$ orbitals (green dots). The situation of the conduction band is the opposite. However, previous studies of the RHF GaBi-X$_2$ monolayer showed that the orbital contributions of the bands close to the Fermi level are mainly from the $s$, $p_x$, and $p_y$ without $p_z$ [71-78]. We may conclude that the $p_z$ orbital contributions are due to the partial hybridization between the $p_z$ orbitals and the X atomic orbitals, which comes from the deviations along the x direction between the X atoms and the Ga/Bi atoms. It is noteworthy that the spin degeneracy is lifted due to the asymmetric geometric structure. Such spin-splitting effect has also been found in III-Bi [24], hydrogenated/functionalized III-Bi [71-76], and g-TlA (A= N, P, As and Sb) [53] monolayers. This lifting of spin degeneracy is due to spin-orbit interaction and results in terms linear in electron wave vector ***k*** in the effective Hamiltonian. The origin of these terms linear in low-dimensional systems is structure and bulk inversion asymmetry which lead to Rashba and Dresselhaus spin-orbit terms in the Hamiltonian, respectively [99-102]. It is well known that spin-splitting of Rashba states in a two-dimensional electron system provides a promising mechanism for spin manipulation that is needed in spintronics applications [103].

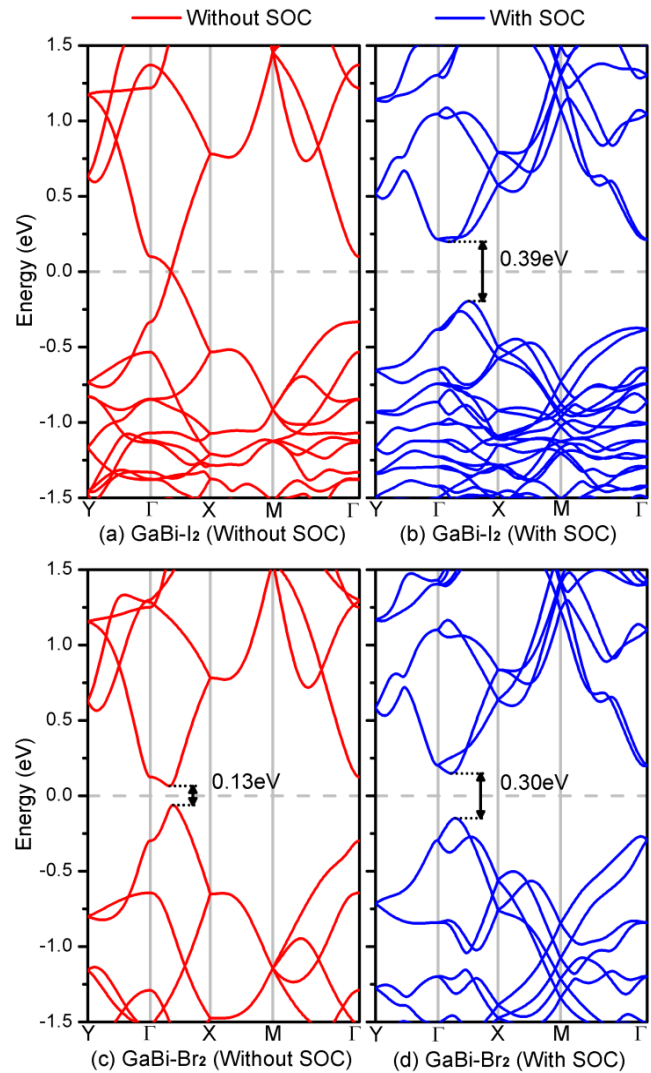

**Figure 3**. Band structures of the DHF GaBi-$I_2$ and GaBi-$Br_2$ monolayers with and without SOC from DFT calculations. (a) GaBi-$I_2$ without SOC, (b) GaBi-$I_2$ with SOC, (c) GaBi-$Br_2$ without SOC, and (d) GaBi-$Br_2$ with SOC.

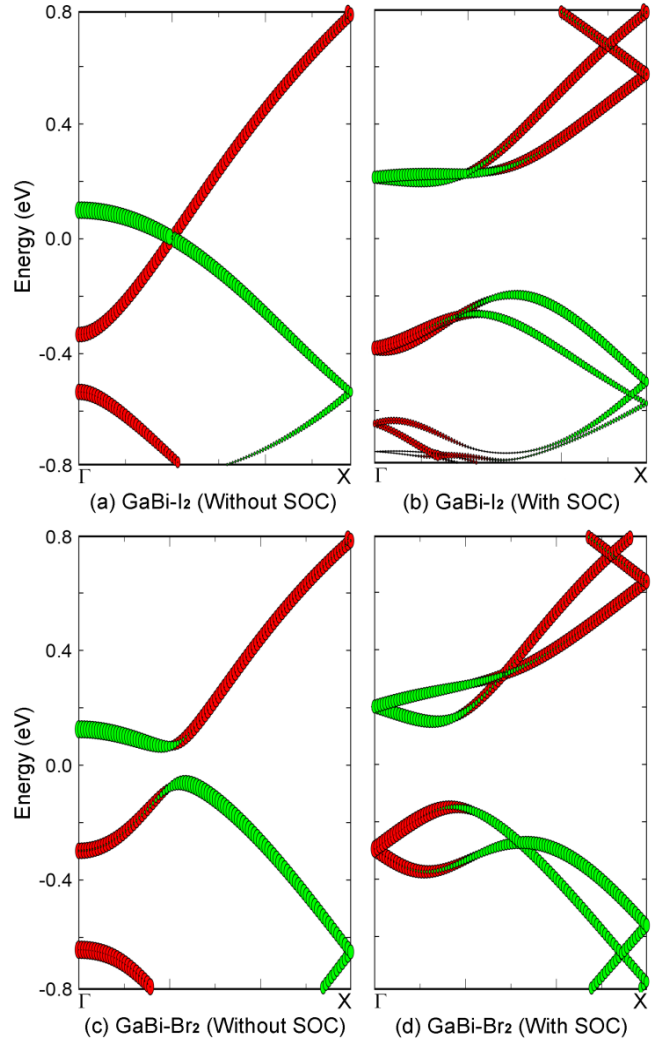

**Figure 4**. Orbital-projected band structures close to the Fermi level with and without SOC along the Γ-X line. (a) GaBi-I$_2$ without SOC, (b) GaBi-I$_2$ with SOC, (c) GaBi-Br$_2$ without SOC, and (d) GaBi-Br$_2$ with SOC. The red dots represent the contributions from the $s$, $p_x$, and $p_z$ atomic orbitals of the Ga and Bi atoms and the green dots represent contributions from the $p_y$ atomic orbitals of the Ga and Bi atoms.

In contrast to the gapless band structure of the DHF GaBi-I$_2$ monolayer, the band structure without SOC of the DHF GaBi-Br$_2$ monolayer shows semiconducting behavior. The indirect band gap is 0.13 eV without SOC (Figure 3(c)) while it becomes 0.30 eV with SOC (Figure 3(d)), which is less than that of the RHF GaBi-Br$_2$ (0.628 eV) [75]. From the corresponding orbital-projected band structures close to the

Fermi level (Figure 4(c) and (d)), we find that those bands result from the *s* and *p* atomic orbitals of the Ga and Bi atoms. They are similar to the band structure with SOC of the DHF GaBi-I$_2$, which can also be divided into two parts. The SOC band structure also shows obvious spin-splitting of Rashba states. The band structures without and with SOC of the DHF GaBi-Cl$_2$ monolayer are shown in Figure S2(a) and (c) (Supplementary Information) and are similar to those of the DHF GaBi-Br$_2$. An indirect band gap of 0.22 eV (0.17eV) is found without (with) SOC. The band gap with SOC is much less than that of the RHF GaBi-Cl$_2$ monolayer, which was found to be 0.645eV [75].

**Topologically Nontrivial Properties**

Similar to the band structures of graphene (silicene, germanene, and stanene) [38, 41, 42] and other 2D topological insulators, the DHF GaBi-I$_2$ monolayer also has a Dirac cone in its band structure without SOC, which is an important sign of a QSH insulator [41, 42, 104, 105] and a large SOC bulk gap in its band structure with SOC. To confirm the topologically nontrivial property of the bulk gap, we calculated its Z$_2$ topological invariant (see Table 1), which is equal to 1. This indicates that the DHF GaBi-I$_2$ monolayer is a 2D topological insulator. For the DHF GaBi-Br$_2$ and GaBi-Cl$_2$ monolayers, we also calculated the Z$_2$ topological invariant (see Table 1). We find that the Z$_2$ topological invariant is 1 in both cases, although there are no Dirac cones in their band structures without SOC. To find the Dirac cone, we applied different SOC strengths ($\lambda_{SOC}$) [104, 105] and calculated the corresponding band gaps of the DHF GaBi-Br$_2$ (Figure 5(a)). The band gap decreases with increasing $\lambda_{SOC}$ at the beginning. When $\lambda_{SOC}$ reaches 0.24, a gapless state can be found. Then the band gap increases with increasing $\lambda_{SOC}$. We plot the band structure for $\lambda_{SOC}$ = 0.24 (Figure 5(b)) and a distorted Dirac cone can be seen. From its corresponding orbital-projected band structure close to the Fermi level along the Γ-X line, it is seen that the crossing bands forming the distorted Dirac cone arise from the *s* and *p* atomic orbitals of the Ga and Bi atoms, similar to the case of the DHF GaBi-I$_2$ monolayer. For the DHF GaBi-Cl$_2$ monolayer, we find a similar distorted Dirac cone when $\lambda_{SOC}$ is equal to 0.43 (Supplementary Information, Figure S2(b)).

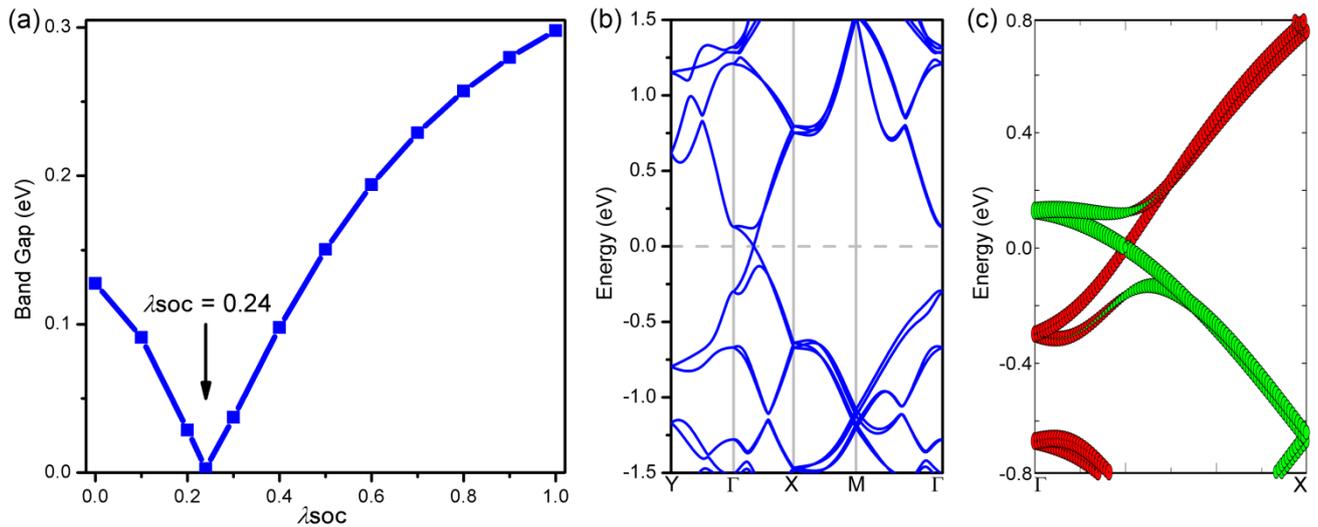

**Figure 5**. (a) Variation of the DHF GaBi-Br$_2$ band gap as a function of SOC strength $\lambda_{SOC}$. (b) Band structure of the DHF GaBi-Br$_2$ for $\lambda_{SOC} = 0.24$. (c) Orbital-projected band structure for $\lambda_{SOC}=0.24$ close to the Fermi level. The red dots represent the contributions from the *s*, $p_x$, and $p_z$ atomic orbitals of the Ga and Bi atoms and the green dots represent contributions from the $p_y$ atomic orbitals of the Ga and Bi atoms.

Although iodization, bromination, and chlorination of GaBi lead to qualitatively similar bands close to the Fermi level that arise from the atomic orbitals of the Ga and Bi atoms, their band gaps are totally different. We conclude that the different functionalizd elements lead to different build-in electric fields, which can influence the band gaps [74]. For the DHF GaBi-X$_2$ monolayer, we can only find the Dirac cone for a particular $\lambda_{SOC}$ and if we tune the $\lambda_{SOC}$ or change the calculation method, the Dirac cone will disappear. It can be seen that the Dirac cone of the DHF GaBi-I$_2$ disappears for $\lambda_{SOC} = 0$, when we use the more sophisticated Heyd-Scuseria-Ernzerhof (HSE06) [106, 107] hybrid functional method (Supplementary Information, Figure S3). With HSE, the band gap including SOC is 0.47 eV (Figure S3(b)), which is a little larger than the 0.39 eV of the PBE calculations. For the HSE without SOC, a small band gap (0.05 eV) is opened (Figure S3(a)), which is different from the Dirac cone of the PBE results. Similar to the band structures of the GaBi-Br$_2$ and GaBi-Cl$_2$ monolayers, the appearance of the DHF GaBi-I$_2$ Dirac cone in the HSE calculations show up for a particular $\lambda_{SOC}$ between 0 and 1[104]. It is certain that we can always find Dirac cones for the three DHF GaBi-X$_2$ monolayers when varying $\lambda_{SOC}$ from 0 to 1, but we can't find a Dirac cone for the DHF GaBi-F$_2$ monolayer. Its band gap ranges from 0.32 eV ($\lambda_{SOC} = 0$) to 0.03 eV ($\lambda_{SOC} = 1$) (Supplementary Information, Figure S4) and we can't find the gapless state in the process. Corresponding to the absence of the Dirac cone, its Z$_2$ topological invariant is equal to 0, which indicates that the DHF GaBi-F$_2$ monolayer is a trivial insulator.

Besides the nonzero Z$_2$ topological invariant, the existence of gapless edge states is another prominent feature of QSH insulators. Since the DHF GaBi-Br$_2$ and GaBi-Cl$_2$ monolayers show a similar band structure, without loss of generality, we only considered the DHF GaBi-I$_2$ and GaBi-Br$_2$ nanoribbons with armchair edges. The nanoribbon structure is shown in Figure 6(a) and their widths (*W*) are wide enough to avoid interactions between the edge states from the two sides. The band structures of the DHF GaBi-I$_2$ and GaBi-Br$_2$ nanoribbons are shown in Figure 6(b) and (c), respectively. We can see that the gapless edge states (Dirac point) appear in the bulk gap and the bands cross linearly at the Γ point, demonstrating the topologically nontrivial nature of these 2D materials.

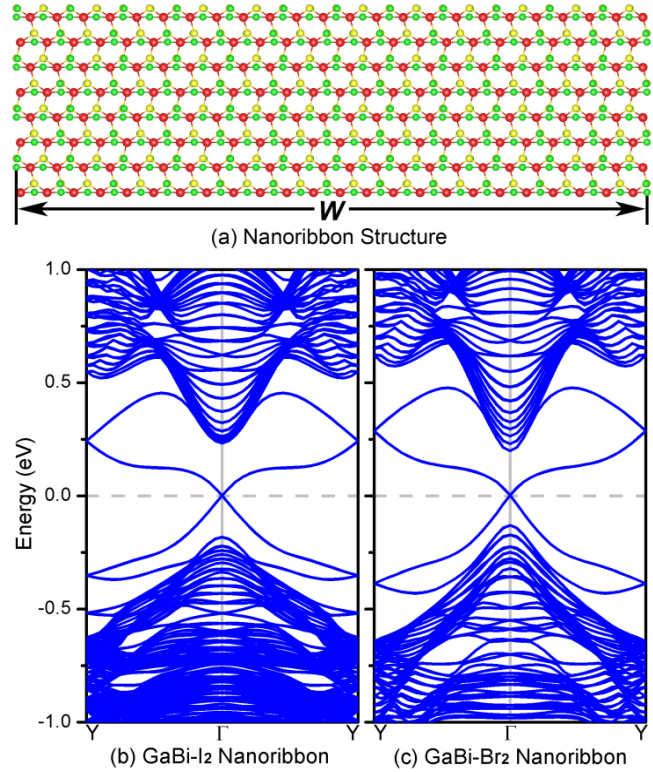

**Figure 6**. (a) Top view of the nanoribbon structure. The width of the nanoribbon is labeled as *W*. Electronic band structures of the GaBi-$I_2$ nanoribbon with *W*=10.2 nm (b) and the GaBi-$Br_2$ nanoribbon with *W*=9.9 nm (c).

**Conclusion**

In summary, using first-principles calculations, we predict that the DHF GaBi-$X_2$ (X=I, Br, Cl) monolayers are QSH insulators with: 1) a novel framework structure, 2) high stability, 3) sizeable nontrivial bulk gap, and 4) exhibiting the Rashba effect. The DHF structure is robust against different types of functionalization and is more favorable in energy than the RHF structure. The largest nontrivial bulk gap can reach 0.47 eV, which is beneficial for achieving the room-temperature QSH effect. Corresponding to the nonzero $Z_2$ topological invariant, band calculations show that their distorted Dirac cones will appear in the process of tuning the SOC strength. The nanoribbon edge states with Dirac spectrum are very promising for future spintronics device applications.

**Supplementary Information**

There are four figures in Supplementary Information. Figure S1 is the Snapshot taken from MD simulations for a 2×2 supercell of the DHF GaBi-I$_2$ monolayer at temperature of 300K after 4ps. Figure S2 is the band structures of the DHF GaBi-Cl$_2$ monolayer for $\lambda_{SOC} = 0$, $\lambda_{SOC} = 0.43$, and $\lambda_{SOC} = 1$. Figure S3 is the band structures of the DHF GaBi-I$_2$ monolayer with and without SOC from the HSE calculations. Figure S4 shows the band structures of the DHF GaBi-F$_2$ monolayer with and without SOC.

**Acknowledgements**

This work was supported by the Fonds Wetenschappelijk Onderzoek (FWO-Vl). Computational resources were provided by HPC infrastructure of the University of Antwerp (CalcUA) a division of the Flemish Supercomputer Center (VSC), which is funded by the Hercules foundation and the Flemish Government department EWI.

**References**

[1] Novoselov, K. S.; Geim, A. K.; Morozov, S. V.; Jiang, D.; Zhang, Y.; Dubonos, S. V.; Grigorieva, I. V.; Firsov, A. A. Electric Field Effect in Atomically Thin Carbon Films. *Science* **2004**, *306*, 666-669.

[2] Guzmán-Verri, G. G.; Lew Yan Voon, L. C. Electronic structure of silicon-based nanostructures. *Phys. Rev. B* **2007**, *76*, 075131.

[3] Jamgotchian, H.; Colignon, Y.; Hamzaoui, N.; Ealet, B.; Hoarau, J. Y.; Aufray, B.; Bibérian, J. P. Growth of silicene layers on Ag(111): unexpected effect of the substrate temperature. *J. Phys.: Condens. Matter* **2012**, *24*, 172001.

[4] Lalmi, B.; Oughaddou, H.; Enriquez, H.; Kara, A.; Vizzini, S.; Ealet, B; Aufray, B. Epitaxial growth of a silicene sheet. *Appl. Phys. Lett.* **2010**, *97*, 223109.

[5] Feng, B.; Ding, Z.; Meng, S.; Yao, Y.; He, X.; Cheng, P.; Chen, L.; Wu, K. Evidence of Silicene in Honeycomb Structures of Silicon on Ag(111). *Nano Lett.* **2012**, *12*, 3507-3511.

[6] Vogt, P.; De Padova, P.; Quaresima, C.; Avila, J.; Frantzeskakis, E.; Asensio, M. C.; Resta, A.; Ealet, B.; Le Lay, G. Silicene: Compelling Experimental Evidence for Graphenelike Two-Dimensional Silicon. *Phys. Rev. Lett.* **2012**, *108*, 155501.

[7] Lin, C. L.; Arafune, R.; Kawahara, K.; Tsukahara, N.; Minamitani, E.; Kim, Y.; Takagi, N.; Kawai, M. Structure of Silicene Grown on Ag(111). *Appl. Phys. Express* **2012**, *5*, 045802.

[8] Chiappe, D.; Grazianetti, C.; Tallarida, G.; Fanciulli, M.; Molle, A. Local Electronic Properties of Corrugated Silicene Phases. *Adv. Mater.* **2012**, *24*, 5088-5093.

[9] Arafune, R.; Lin, C. L.; Kawahara, K.; Tsukahara, N.; Minamitani, E.; Kim, Y.; Takagi, N.; Kawai, M. Structural transition of silicene on Ag(111). *Surf. Sci.* **2013**, *608*, 297-300.

[10] Feng, B.; Li, H.; Liu, C. C.; Shao, T. N.; Cheng, P.; Yao, Y.; Meng, S.; Chen, L.; Wu, K. Observation of Dirac Cone Warping and Chirality Effects in Silicene. *ACS Nano* **2013**, *7*, 9049-9054.

[11] Meng, L.; Wang, Y.; Zhang, L.; Du, S.; Wu, R.; Li, L.; Zhang, Y.; Li, G.; Zhou, H.; Hofer, W. A. *et al* Buckled Silicene Formation on Ir(111). *Nano Lett.* **2013**, *13*, 685-690.

[12] Fleurence, A.; Friedlein, R.; Ozaki, T.; Kawai, H.; Wang, Y.; Yamada-Takamura, Y. Experimental Evidence for Epitaxial Silicene on Diboride Thin Films. *Phys. Rev. Lett.* **2012**, *108*, 245501.

[13] Chiappe, D.; Scalise, E.; Cinquanta, E.; Grazianetti, C.; van den Broek, B.; Fanciulli, M.; Houssa, M.; Molle, A. Two-Dimensional Si Nanosheets with Local Hexagonal Structure on a MoS$_2$ Surface. *Adv. Mater.* **2014**, *26*, 2096-2101.

[14] Dávila, M. E.; Xian, L.; Cahangirov, S.; Rubio, A.; Le Lay, G. Germanene: a novel two-dimensional germanium allotrope akin to graphene and silicene. *New J. Phys.* **2014**, *16*, 095002.

[15] Bianco, E.; Butler, S.; Jiang, S.; Restrepo, O. D.; Windl, W.; Goldberger, J. E. Stability and Exfoliation of Germanane: A Germanium Graphane Analogue. *ACS Nano* **2013**, *7*, 4414-4421.

[16] Jiang, S.; Butler, S.; Bianco, E.; Restrepo, O. D.; Windl, W.; Goldberger, J. E. Improving the stability and optical properties of germanane via one-step covalent methyl-termination. *Nat. Commun.* **2014**, *5*, 3389.

[17] Derivaz, M.; Dentel, D.; Stephan, R.; Hanf, M. C.; Mehdaoui, A.; Sonnet, P.; Pirri, C. Continuous germanene layer on Al(111). *Nano Lett.* **2015**, *15*, 2510-2516.

[18] Zhu, F. F.; Chen, W. J.; Xu, Y.; Gao, C. L.; Guan, D. D.; Liu, C. H.; Qian, D.; Zhang, S. C.; Jia, J. F. Epitaxial growth of two-dimensional stanene. *Nature Mater.* **2015**, *14*, 1020-1025.

[19] Radisavljevic, B.; Radenovic, A.; Brivio, J.; Giacometti, V.; Kis, A. Single-layer MoS$_2$ transistors. *Nature Nanotech.* **2011**, *6*, 147-150.

[20] Mak, K. F.; Lee, C.; Hone, J.; Shan, J.; Heinz, T. F. Atomically Thin MoS$_2$: A New Direct-Gap Semiconductor. *Phys. Rev. Lett.* **2010**, *105*, 136805.

[21] Zhang, Y.; Zhang, Y.; Ji, Q.; Ju, J.; Yuan, H.; Shi, J.; Gao, T.; Ma, D.; Liu, M.; Chen, Y. *et al* Controlled Growth of High-Quality Monolayer WS$_2$ Layers on Sapphire and Imaging Its Grain Boundary. *ACS Nano* **2013**, *7*, 8963-8971.

[22] Zhou, H.; Zhao, M.; Zhang, X.; Dong, W.; Wang, X.; Bu, H.; Wang, A. First-principles prediction of a new Dirac-fermion material: silicon germanide monolayer. *J. Phys.: Condens. Matter* **2013**, *25*, 395501.

[23] Li, J.; He, C.; Meng, L.; Xiao, H.; Tang, C.; Wei, X.; Kim, J.; Kioussis, N.; Malcolm Stocks, G.; Zhong, J. Two-dimensional topological insulators with tunable band gaps: Single-layer HgTe and HgSe. *Sci. Rep.* **2015**, *5*, 14115.

[24] Chuang, F. C.; Yao, L. Z.; Huang, Z. Q.; Liu, Y. T.; Hsu, C. H.; Das, T.; Lin, H.; Bansil, A. Prediction of Large-Gap Two-Dimensional Topological Insulators Consisting of Bilayers of Group III Elements with Bi. *Nano Lett.* **2014**, *14*, 2505-2508.

[25] Wang, Z. F.; Su, N.; Liu, F. Prediction of a Two-Dimensional Organic Topological Insulator. *Nano Lett.* **2013**, *13*, 2842-2845.

[26] Wang, Z. F.; Liu, Z.; Liu, F. Organic topological insulators in organometallic lattices. *Nat. Commun.* **2013**, *4*, 1471.

[27] Wang, Z. F.; Liu, Z.; Liu, F. Quantum Anomalous Hall Effect in 2D Organic Topological Insulators. *Phys. Rev. Lett.* **2013**, *110*, 196801.

[28] Liu, Z.; Wang, Z. F.; Mei, J. W.; Wu, Y. S.; Liu, F. Flat Chern Band in a Two-Dimensional Organometallic Framework. *Phys. Rev. Lett.* **2013**, *110*, 106804.

[29] Qi, X. L; Zhang, S. C. Topological insulators and superconductors. *Rev. Mod. Phys.* **2011**, *83*, 1057-1110.

[30] Hasan, M. Z.; Kane, C. L. Colloquium: Topological insulators. *Rev. Mod. Phys.* **2010**, *82*, 3045-3067.

[31] Bernevig, B. A.; Zhang, S. C. Quantum Spin Hall Effect. *Phys. Rev. Lett.* **2006**, *96*, 106802/.

[32] Bernevig, B. A.; Hughes, T. L.; Zhang, S. C. Quantum Spin Hall Effect and Topological Phase Transition in HgTe Quantum Wells. *Science* **2006**, *314*, 1757-1761.

[33] König, M.; Wiedmann, S.; Brüne, C.; Roth, A.; Buhmann, H.; Molenkamp, L. W.; Qi, X. L.; Zhang, S. C. Quantum Spin Hall Insulator State in HgTe Quantum Wells. *Science* **2007**, *318*, 766-770.

[34] Knez, I.; Du, R. R.; Sullivan, G. Evidence for Helical Edge Modes in Inverted InAs/GaSb QuantumWells. *Phys. Rev. Lett.* **2011**, *107*, 136603.

[35] Ren, Y.; Qiao, Z.; Niu, Q. Topological phases in two-dimensional materials: a review. *Rep. Prog. Phys.* **2016**, *79*, 066501.

[36] Takayama, A.; Sato, T.; Souma, S.; Oguchi, T.; Takahashi, T. One-Dimensional Edge States with Giant Spin Splitting in a Bismuth Thin Film. *Phys. Rev. Lett.* **2015**, *114*, 066402.

[37] Drozdov, I. K.; Alexandradinata, A.; Jeon, S.; Nadj-Perge, S.; Ji, H.; Cava, R. J.; Bernevig, B.; Yazdani, A. One-dimensional topological edge states of bismuth bilayers. *Nature Phys.* **2014**, *10* 664-669

[38] Kane, C. L.; Mele, E. J. Quantum Spin Hall Effect in Graphene. *Phys. Rev. Lett.* **2005**, *95*, 226801.

[39] Kou, L.; Yan, B.; Hu, F.; Wu, S. C.; Wehling, T. O.; Felser, C.; Chen, C.; Frauenheim, T.

Graphene-Based Topological Insulator with an Intrinsic Bulk Band Gap above Room Temperature. *Nano Lett.* **2013**, *13*, 6251-6255.

[40] Kou, L.; Wu, S. C.; Felser, C.; Frauenheim, T.; Chen, C.; Yan, B. Robust 2D Topological Insulators in van der Waals Heterostructures. *ACS Nano* **2014**, *8*, 10448-10454.

[41] Liu, C. C.; Feng, W.; Yao, Y. Quantum Spin Hall Effect in Silicene and Two-Dimensional Germanium. *Phys. Rev. Lett.* **2011**, *107*, 076802.

[42] Liu, C. C.; Jiang, H.; Yao, Y. Low-energy effective Hamiltonian involving spin-orbit coupling in silicene and two-dimensional germanium and tin. *Phys. Rev. B* **2011**, *84*, 195430.

[43] Tang, P.; Chen, P.; Cao, W.; Huang, H.; Cahangirov, S.; Xian, L.; Xu, Y.; Zhang, S. C.; Duan, W.; Rubio, A. Stable two-dimensional dumbbell stanene: A quantum spin Hall insulator. *Phys. Rev. B* **2014**, *90*, 121408(R).

[44] Chen, X.; Li, L.; Zhao, M. Hydrogenation-induced large-gap quantum-spin-Hall insulator states in a germanium–tin dumbbell structure. *RSC Adv.* **2015**, *5*, 72462-72468.

[45] Chen, X.; Li, L.; Zhao, M. Dumbbell stanane: a large-gap quantum spin hall insulator. *Phys. Chem. Chem. Phys.* **2015**, *17*, 16624-16629

[46] Nie, S. M.; Song, Z.; Weng, H.; Fang, Z. Quantum spin Hall effect in two-dimensional transition-metal dichalcogenide haeckelites. *Phys. Rev. B* **2015**, *91*, 235434.

[47] Li, W.; Guo, M.; Zhang, G.; Zhang, Y. W. Gapless $MoS_2$ allotrope possessing both massless Dirac and heavy fermions. *Phys. Rev. B* **2014**, *89*, 205402.

[48] Ma, Y.; Kou, L.; Li, X.; Dai, Y.; Heine, T. Two-dimensional transition metal dichalcogenides with a hexagonal lattice: Room-temperature quantum spin Hall insulators. *Phys. Rev. B* **2016**, *93*, 035442.

[49] Ma, Y.; Kou, L.; Li, X.; Dai, Y.; Smith, S. C.; Heine, T. Quantum spin Hall effect and topological

phase transition in two-dimensional square transition-metal dichalcogenides. *Phys. Rev. B* **2015**, *92*, 085427.

[50] Sun, Y.; Felser, C.; Yan, B. Graphene-like Dirac states and quantum spin Hall insulators in square-octagonal $MX_2$ (M =Mo; W; X =S; Se; Te) isomers. *Phys. Rev. B* **2015**, *92*, 165421.

[51] Liu, P. F.; Zhou, L.; Frauenheim, T.; Wu, L. M. New quantum spin Hall insulator in two-dimensional $MoS_2$ with periodically distributed pores. *Nanoscale* **2016**, *8*, 4915-4921.

[52] Qian, X.; Liu, J.; Fu, L.; Li, J. Quantum spin Hall effect in two-dimensional transition metal dichalcogenides. *Science* **2014**, *346*, 1344-1347.

[53] Li, X.; Dai, Y.; Ma, Y.; Wei, W,; Yu, L.; Huang, B. Prediction of large-gap quantum spin hall insulator and Rashba-Dresselhaus effect in two-dimensional g-TlA (A = N; P; As; and Sb) monolayer films. *Nano Research* **2015**, *8*, 2954-2962.

[54] Zhou, L.; Kou, L.; Sun, Y.; Felser, C.; Hu, F.; Shan, G.; Smith, S. C.; Yan, B.; Frauenheim, T. New Family of Quantum Spin Hall Insulators in Two-dimensional Transition-Metal Halide with Large Nontrivial Band Gaps. *Nano Lett.* **2015**, *15*, 7867-7872.

[55] Liu, Z.; Liu, C. X.; Wu, Y. S.; Duan, W. H.; Liu, F.; Wu, J. Stable Nontrivial $Z_2$ Topology in Ultrathin Bi (111) Films: A First-Principles Study. *Phys. Rev. Lett.* **2011**, *107*, 136805.

[56] Weng, H.; Dai, X.; Fang, Z. Transition-Metal Pentatelluride $ZrTe_5$ and $HfTe_5$: A Paradigm for Large-Gap Quantum Spin Hall Insulators. *Phys. Rev. X* **2014**, *4*, 011002

[57] Kou, L.; Ma, Y.; Yan, B.; Tan, X.; Chen, C.; Smith, S. C. Encapsulated Silicene: A Robust Large-Gap Topological Insulator. *ACS Appl. Mater. Interfaces* **2015**, *7*, 19226-19233

[58] Liu, Q.; Zhang, X.; Abdalla, L. B.; Fazzio, A.; Zunger, A. Switching a Normal Insulator into a Topological Insulator via Electric Field with Application to Phosphorene. *Nano Lett.* **2015**, *15*

1222-1228

[59] Song, Z.; Liu, C. C.; Yang, J.; Han, J.; Ye, M.; Fu, B.; Yang, Y.; Niu, Q.; Lu, J.; Yao, Y. Quantum spin Hall insulators and quantum valley Hall insulators of BiX/SbX (X=H; F; Cl and Br) monolayers with a record bulk band gap. *NPG Asia Materials* **2014**, *6*, e147.

[60] Liu, C. C.; Guan, S.; Song, Z.; Yang, S. A.; Yang, J.; Yao, Y. Low-energy effective Hamiltonian for giant-gap quantum spin Hall insulators in honeycomb X-hydride/halide (X=N–Bi) monolayers. *Phys. Rev. B* **2014**, *90*, 085431

[61] Xu, Y.; Yan, B.; Zhang, H. J.; Wang, J.; Xu, G.; Tang, P.; Duan, W.; Zhang, S. C. Large-Gap Quantum Spin Hall Insulators in Tin Films. *Phys. Rev. Lett.* **2013**, *111*, 136804.

[62] Si, C.; Liu, J.; Xu, Y.; Wu, J.; Gu, B. L.; Duan, W. Functionalized germanene as a prototype of large-gap two-dimensional topological insulators. *Phys. Rev. B* **2014**, *89*, 115429.

[63] Zhou, J. J.; Feng, W.; Liu, C. C.; Guan, S.; Yao, Y. Large-Gap Quantum Spin Hall Insulator in Single Layer Bismuth Monobromide $Bi_4Br_4$. *Nano Lett.* **2014**, *14*, 4767-4771.

[64] Luo, W.; Xiang, H. Room Temperature Quantum Spin Hall Insulators with a Buckled Square Lattice. *Nano Lett.* **2015**, *15*, 3230-3235

[65] Ma, Y.; Dai, Y.; Kou, L.; Frauenheim, T.; Heine, T. Robust Two-Dimensional Topological Insulators in Methyl-Functionalized Bismuth; Antimony; and Lead Bilayer Films. *Nano Lett.* **2015**, *15*, 1083-1089.

[66] Wang, Y. P.; Ji, W. X.; Zhang, C. W.; Li, P.; Li, F.; Ren, M. J.; Chen, X. L.; Yuan, M.; Wang, P. J. Controllable band structure and topological phase transition in two-dimensional hydrogenated arsenene. *Sci. Rep.* **2016**, *6*, 20342.

[67] Zhao, H.; Zhang, C. W.; Ji, W. X.; Zhang, R. W.; Li, S. S.; Yan, S. S.; Zhang, B. M.; Li, P.; Wang, P.

J. Unexpected Giant-Gap Quantum Spin Hall Insulator in Chemically Decorated Plumbene Monolayer. *Sci. Rep.* **2016**, *6*, 20152.

[68] Zhang, R. W.; Zhang, C. W.; Ji, W. X.; Li, S. S.; Yan, S. S.; Hu, S. J.; Li, P.; Wang, P. J.; Li, F. Room Temperature Quantum Spin Hall Insulator in Ethynyl-Derivative Functionalized Stanene Films. *Sci. Rep.* **2016**, *6*, 18879.

[69] Ji, W. X.; Zhang, C. W.; Ding, M.; Li, P.; Li, F.; Ren, M. J.; Wang, P. J.; Hu, S. J.; Yan, S. S. Stanene cyanide: a novel candidate of Quantum Spin Hall insulator at high temperature. *Sci. Rep.* **2015**, *5*, 18604.

[70] Ma, Y.; Dai, Y.; Wei, W.; Huang, B.; Whangbo. M. H. Strain-induced quantum spin Hall effect in methyl-substituted germanane GeCH$_3$. *Sci. Rep.* **2014**, *4*, 7297.

[71] Crisostomo, C. P.; Yao, L. Z.; Huang, Z. Q.; Hsu, C. H.; Chuang, F. C.; Lin, H.; Albao, M. A.; Bansil, A. Robust Large Gap Two-Dimensional Topological Insulators in Hydrogenated III−V Buckled Honeycombs. *Nano Lett.* **2015**, *15*, 6568-6574.

[72] Freitas, R. R. Q.; Rivelino, R.; de Brito Mota, F.; de Castilho, C. M. C.; Kakanakova-Georgieva, A.; Gueorguiev, G. K. Topological Insulating Phases in Two-Dimensional Bismuth-Containing Single Layers Preserved by Hydrogenation. *J. Phys. Chem. C* **2015**, *119*, 23599-23606.

[73] Ma, Y.; Li, X.; Kou, L.; Yan, B.; Niu, C.; Dai, Y.; Heine, T. Two-dimensional inversion-asymmetric topological insulators in functionalized III-Bi bilayers. *Phys. Rev. B* **2015**, *91*, 235306.

[74] Li, L.; Zhang, X.; Chen, X.; Zhao, M. Giant Topological Nontrivial Band Gaps in Chloridized Gallium Bismuthide. *Nano Lett.* **2015**, *15*. 1296-1301.

[75] Freitas. R. R. Q.; de Brito Mota, F.; Rivelino, R.; de Castilho, C. M. C.; Kakanakova-Georgieva, A.; Gueorguiev, G. K. Tuning band inversion symmetry of buckled III-Bi sheets by halogenation.

*Nanotechnology* **2016**, *27*, 055704.

[76] Li, S. S.; Ji, W. X.; Zhang, C. W.; Hu, S. J.; Li, P.; Wang, P. J.; Zhang, B. M.; Cao, C. L. Robust Room-Temperature Quantum Spin Hall Effect in Methyl-functionalized InBi honeycomb film. *Sci. Rep.* **2016**, *6*, 23242.

[77] Zhang, R. W.; Zhang, C. W.; Ji, W. X.; Li, S. S.; Yan, S. S.; Li, P.; Wang, P. J. Functionalized Thallium Antimony Films as Excellent Candidates for Large-Gap Quantum Spin Hall Insulator. *Sci. Rep.* **2016**, *6*, 21351.

[78] Zhao, M.; Chen, X.; Li, L.; Zhang, X. Driving a GaAs film to a large-gap topological insulator by tensile strain. *Sci. Rep.* **2015**, *5*, 8441.

[79] Kresse, G.; Furthmüller, J. Efficient iterative schemes for ab initio total-energy calculations using a plane-wave basis set. *Phys. Rev. B* **1996**, *54*, 11169-11186.

[80] Kresse, G.; Hafner, J. Ab initio molecular dynamics for open-shell transition metals. *Phys. Rev. B* **1993**, *48*, 13115-13118.

[81] Kresse, G.; Joubert, D. From ultrasoft pseudopotentials to the projector augmented-wave method. *Phys. Rev. B* **1999**, *59*, 1758-1775.

[82] Perdew, J. P.; Burke, K.; Ernzerhof, M. Generalized Gradient Approximation Made Simple. *Phys. Rev. Lett.* **1996**, *77*, 3865-3868.

[83] Parlinski, K.; Li, Z. Q.; Kawazoe, Y. First-Principles Determination of the Soft Mode in Cubic $ZrO_2$. *Phys. Rev. Lett.* **1997**, *78*, 4063-4066.

[84] Soluyanov, A. A.; Vanderbilt, D. Computing topological invariants without inversion symmetry. *Phys. Rev. B* **2011**, *83*, 235401.

[85] Fu, L.; Kane, C. L. Time reversal polarization and a $Z_2$ adiabatic spin pump. *Phys. Rev. B* **2006**, *74*,

195312.

[86] Marzari, N.; Vanderbilt, D. Maximally localized generalized Wannier functions for composite energy bands. *Phys. Rev. B* **1997**, *56*, 12847-12865.

[87] Mostofi, A. A.; Yates, J. R.; Lee, Y. S.; Souza, I.; Vanderbilt, D.; Marzari, N. Wannier90: A Tool for Obtaining Maximally-Localised Wannier Functions. *Comput. Phys. Commun.* **2008**, *178*, 685-699.

[88] Henini, M.; Ibáñez, J.; Schmidbauer, M.; Shafi, M.; Novikov, S. V.; Turyanska, L.; Molina, S. I.; Sales, D. L.; Chisholm, M. F.; Misiewicz, J. Molecular beam epitaxy of GaBiAs on ( 311 ) B GaAs substrates. *Appl. Phys. Lett.* **2007**, *91*, 251909.

[89] Francoeur, S.; Seong, M. J.; Mascarenhas, A.; Tixier, S.; Adamcyk, M.; Tiedje, T. Band gap of $GaAs_{1-x}Bi_x$; 0<x<3.6%. *Appl. Phys. Lett.* **2003**, *82*, 3874-3876.

[90] Denisov, N. V.; Alekseev, A. A.; Utas, O. A.; Azatyan, S. G.; Zotov, A. V.; Saranin, A. A. Bismuth–indium two-dimensional compounds on Si(111) surface. *Surf. Sci.* **2016**, *651*, 105-111.

[91] Gruznev, D. V.; Bondarenko, L. V.; Matetskiy, A. V.; Mihalyuk, A. N.; Tupchaya, A. Y.; Utas, O. A.; Eremeev. S. V.; Hsing, C. R.; Chou, J. P.; Wei, C. M. *et al* Synthesis of two-dimensional $Tl_xBi_{1-x}$ compounds and Archimedean encoding of their atomic structure. *Sci. Rep.* **2016**, *6*, 19446.

[92] Ma, F.; Zhou, M.; Jiao, Y.; Gao, G.; Gu, Y.; Billic, A.; Chen, Z.; Du, A. Single Layer Bismuth Iodide: Computational Exploration of Structural; Electrical; Mechanical and Optical Properties. *Sci. Rep.* **2015**, *5*, 17558.

[93] Trotter, J.; Zobel, T. The crystal structure of $SbI_3$ and $BiI_3$. *Z. Kristallogr. – Cryst. Mater.* **1996**, *123*, 67-72.

[94] Morino, Y.; Ukaji, T.; Ito, T. Molecular Structure Determination by Gas Electron Diffraction at High Temperatures. II. Arsenic Triiodide and Gallium Triiodide. *Bull. Chem. Soc. Jpn.* **1996**, *39*, 71-78.

[95] Drake, M. C.; Rosenblatt, G. M. Raman spectroscopy of gaseous GaCl$_3$ and GaI$_3$. *J. Chem. Phys.* **1976**, *65*, 4067-4071.

[96] Saboungi, M. L.; Howe, M. A.; Price, D. L. Structure and dynamics of molten aluminium and gallium trihalides I. Neutron diffraction. *Mol. Phys.* **1993**, *79*, 847-858.

[97] Wang, Z.; Zhou, X. F.; Zhang, X.; Zhu, Q.; Dong, H.; Zhao, M.; Oganov, A. R. Phagraphene: A Low-Energy Graphene Allotrope Composed of 5−6−7 Carbon Rings with Distorted Dirac Cones. *Nano Lett.* **2015**, *15*, 6182-6186.

[98] Zhou, L.; Shi, W.; Sun, Y.; Shao, B.; Felser, C.; Yan, B.; Frauenheim, T. Two-dimensional rectangular tantalum carbide halides TaCX (X= Cl, Br, I): novel large-gap quantum spin Hall insulator. *2D Mater.* **2016**, *3*, 035018.

[99] Giglberger, S.; Golub, L. E.; Bel'kov, V. V.; Danilov, S. N.; Schuh, D.; Gerl, C.; Rohlfing, F.; Stahl, J.; Wegscheider, W.; Weiss, D.; Prettl, W. *et al* Rashba and Dresselhaus spin splittings in semiconductor quantum wells measured by spin photocurrents. *Phys. Rev. B* **2007**, *75*, 035327.

[100] Bychkov, Y. A.; Rashba, É. I. Properties of a 2D electron gas with lifted spectral degeneracy. *JETP Lett.* **1984**, *39*, 78−81.

[101] Lommer, G.; Malcher, F.; Rössler, U. Reduced *g* factor of subband Landau levels in AlGaAs/GaAs heterostructures. *Phys. Rev. B* **1985**, *32*, 6965−6967 (R).

[102] Žutić, I.; Fabian, J.; Das Sarma, S. Spintronics: Fundamentals and applications. *Rev. Mod. Phys.* **2004**, *76*, 323−410.

[103] Ming, W.; Wang, Z. F.; Zhou, M.; Yoon, M.; Liu, F. Formation of Ideal Rashba States on Layered Semiconductor Surfaces Steered by Strain Engineering. *Nano Lett.* **2016**, *16*, 404-409.

[104] Kou, L.; Tan, X.; Ma, Y.; Tahini, H.; Zhou, L.; Sun, Z.; Du, A.; Chen, C.; Smith, S. C. Tetragonal

bismuth bilayer: a stable and robust quantum spin hall insulator. *2D Mater.* **2015**, *2*, 045010.

[105] Huang, B.; Jin, K. H.; Zhuang, H. L.; Zhang, L.; Liu, F. Interface orbital engineering of large-gap topological states: Decorating gold on a Si(111) surface. *Phys. Rev. B* **2016**, *93*, 115117.

[106] Heyd, J.; Scuseria, G. E.; Ernzerhof, M. Hybrid functionals based on a screened Coulomb potential. *J. Chem. Phys.* **2003**, *118*, 8207-8215.

[107] Heyd, J.; Scuseria, G. E.; Ernzerhof, M. Erratum: "Hybrid functionals based on a screened Coulomb potential" [J. Chem. Phys.118; 8207 (2003)]. *J. Chem. Phys.* **2006**, *124*, 219906.